\documentclass[twocolumn,showpacs,prl]{revtex4-1}
\usepackage{graphicx}
\usepackage{color}
\usepackage{amssymb}

\begin{document}
\title{Material evidence of a 38 MeV boson}
\author{Eef van Beveren$^{1}$ and George Rupp$^{2}$}
\affiliation{
$^{1}$Centro de F\'{\i}sica Computacional,
Departamento de F\'{\i}sica, Universidade de Coimbra,
P-3004-516 Coimbra, Portugal\\
$^{2}$Centro de F\'{\i}sica das Interac\c{c}\~{o}es Fundamentais,
Instituto Superior T\'{e}cnico, Technical University of Lisbon,
P-1049-001 Lisboa, Portugal
}
\date{\today}

\begin{abstract}
We present further and more compelling evidence of the existence
of $E(38)$, a light boson that most probably
couples exclusively to quarks and gluons.
Observations presented in a prior paper will be rediscussed for completeness.
\end{abstract}


\maketitle

In a recent paper \cite{ARXIV11021863},
we presented a variety of indications of the possible existence
of a light boson with a mass of about 38 MeV,
henceforth referred to as $E(38)$.
These indications amounted to a series of
low-statistics observations all pointing in the same direction,
which will be discussed here anew, for completeness.
However intriguing, the pieces just did not yet add up
to a full and clear picture in Ref.~\cite{ARXIV11021863}.
In this Letter, we present three more pieces of evidence, one of which
being considerably more conclusive owing to the much higher statistics.
However, the latter data do not confirm the existence
of recurrencies of $E(38)$, as previously \cite{ARXIV11021863}
suggested.

\begin{figure}[ht]
\begin{center}
\begin{tabular}{c}
\includegraphics[width=240pt]{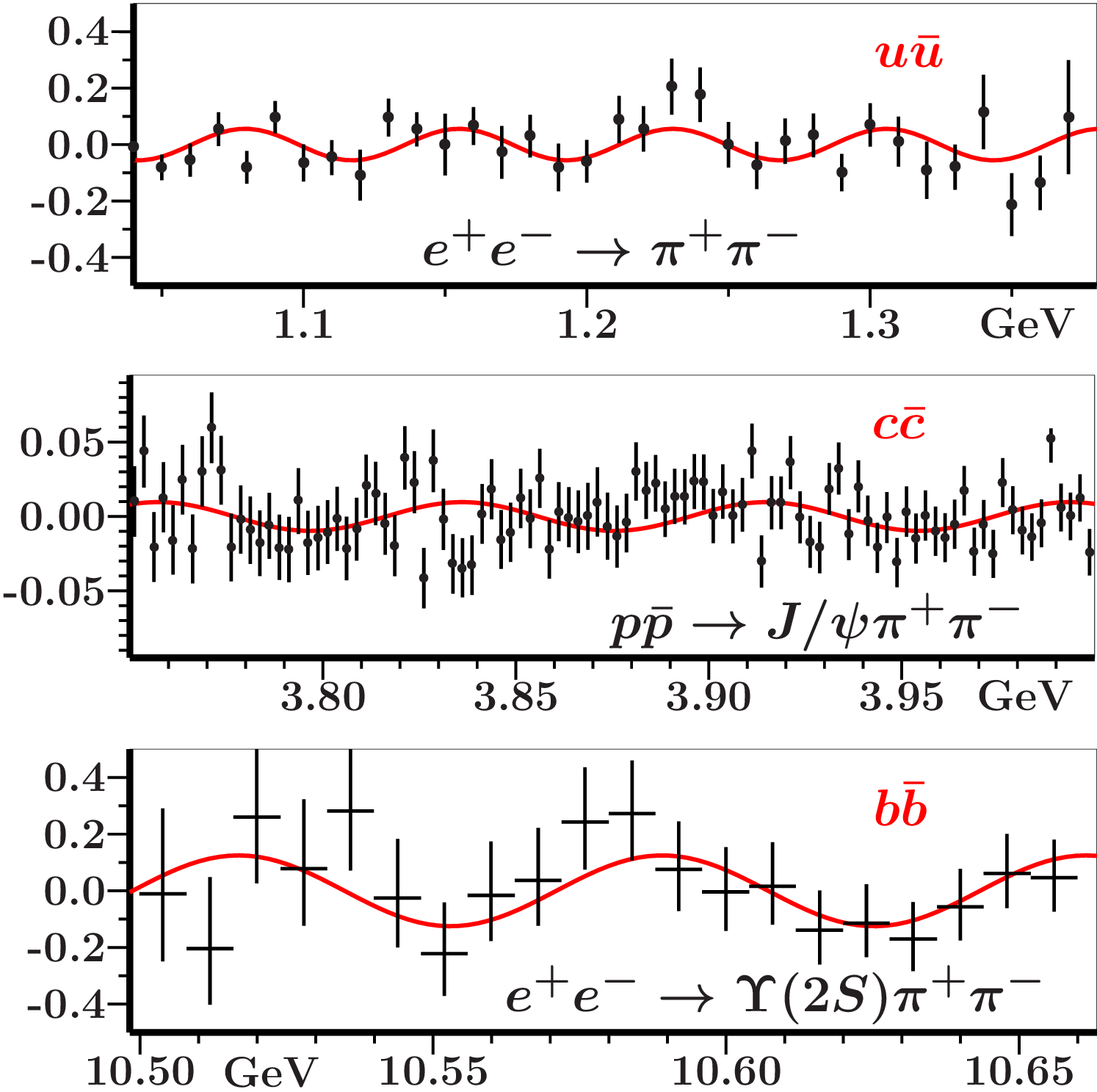}\\ [-10pt]
\end{tabular}
\end{center}
\caption{\small
Fits to the residual data, after subtraction of global fits to:
$e^{+}e^{-}\to\pi^{+}\pi^{-}$ data of the CMD-2 Collaboration
\cite{JETPL82p743}, with a period of 78$\pm$2 MeV
and an amplitude of $\approx$5\% ({\it top});
$p\bar{p}\to J/\psi\pi^{+}\pi^{-}$ data
of the CDF Collaboration \cite{PRL103p152001},
with a period of 79$\pm$5 MeV
and an amplitude of about 0.75\% ({\it middle});
$e^{+}e^{-}\to\Upsilon (2S)\pi^{+}\pi^{-}$ data
of the BABAR Collaboration \cite{PRD78p112002},
with a period of 73$\pm$3 MeV
and an amplitude of some 12.5\% ({\it bottom}).
}
\label{interference}
\end{figure}

In Ref.~\cite{PRD79p111501R},
we made notice of an apparent interference effect
around the $D_{s}^{\ast}\bar{D}_{s}^{\ast}$ threshold
in the invariant-mass distribution
of $e^{+}e^{-}\to J/\psi\pi^{+}\pi^{-}$ events,
which we observed in preliminary radiation data
of the BABAR Collaboration \cite{ARXIV08081543}.
The effect, with a periodicity of about 74 MeV,
could be due to interference
between the typical oscillation frequency of 190 MeV
of the $c\bar{c}$ pair,
as in the model of Refs.~\cite{PRD21p772,PRD27p1527},
and that of the gluon cloud.
Later, in Ref.~\cite{ARXIV10095191},
we reported evidence of small oscillations
in electron-positron and proton-antiproton annihilation data,
with a periodicity of 76$\pm$2 MeV, independent of the beam energy.
The latter observations are summarized in Fig.~\ref{interference}.

Amongst the various scenarios to explain the phenomenon
presented in Ref.~\cite{ARXIV10095191},
one was rather intriguing, namely the postulated existence of
gluonic oscillations, possibly surface oscillations,
with a frequency of about 38 MeV. These would then, upon interfering
with the universal quarkonia frequency $\omega =190$ MeV
\cite{PRD21p772,PRD27p1527},
lead to the observed oscillations.

Here, we will find that the phenomenon
is most likely to be associated with the
interquark exchange of a boson with a mass of about 38 MeV.
Moreover, from the fact that the observed oscillations
are more intense for bottomonium than for light quarks,
we assume that the coupling of this light boson to quarks
increases with the quark mass.
This seems to correspond well to the scalar particle
of the model of Ref.~\cite{NCA80p401},
and to the enigmatic mass parameter related to
the $^{3\!}P_{0}$ pair-creation mechanism \cite{PRD21p772}.

In Ref.~\cite{PRD78p112002}, the BABAR Collaboration
presented an analysis of data on
$e^{+}e^{-}$ $\to$ $\pi^{+}\pi^{-}
\Upsilon\left( 1,2\,{}^{3\!}S_{1}\right)$
$\to$ $\pi^{+}\pi^{-}\ell^{+}\ell^{-}$
($\ell =e$ and $\ell =\mu$),
with the aim to study hadronic transitions between
$b\bar{b}$ excitations
and the $\Upsilon\left( 1\,{}^{3\!}S_{1}\right)$
and $\Upsilon\left( 2\,{}^{3\!}S_{1}\right)$,
based on 347.5 fb$^{-1}$ of data
taken with the BABAR detector at the PEP-II storage rings.

\begin{figure}[htpb]
\begin{center}
\begin{tabular}{c}
\includegraphics[width=240pt]{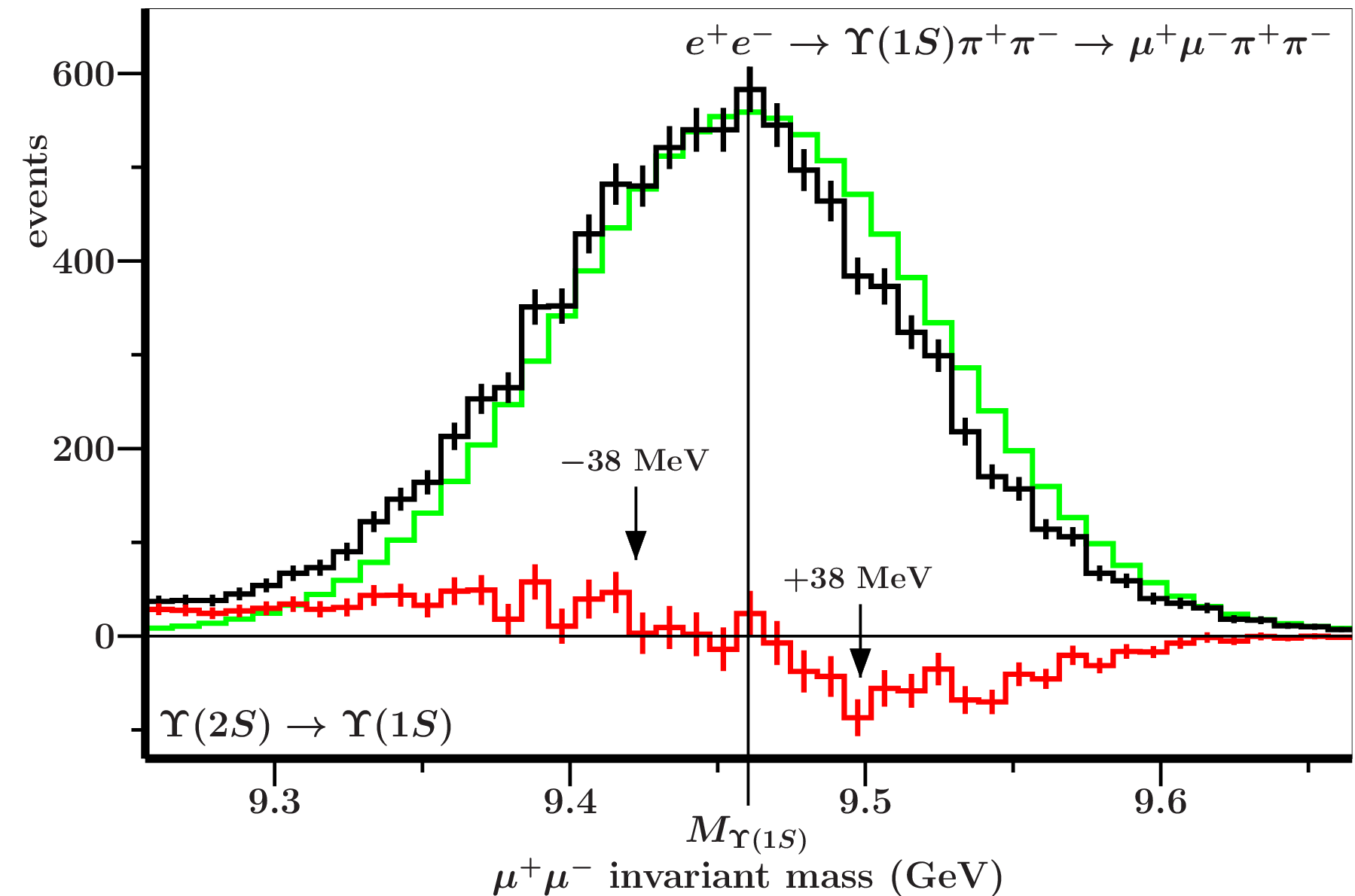}\\ [-10pt]
\end{tabular}
\end{center}
\caption{\small
Invariant $\mu^{+}\mu^{-}$ mass distribution for events
identified as stemming from the reaction
$e^{+}e^{-}$ $\to$ $\Upsilon\left( 2\,{}^{3\!}S_{1}\right)$ $\to$
$\pi^{+}\pi^{-}\Upsilon\left( 1\,{}^{3\!}S_{1}\right)$
$\to$ $\pi^{+}\pi^{-}\mu^{+}\mu^{-}$.
Data (black) are taken from Ref.~\cite{PRD78p112002}.
The bin size equals 9 MeV.
Statistical errors are shown by vertical bars.
The vertical line indicates
$M_{\mu^{+}\mu^{-}}=M_{\Upsilon\left( 1\,{}^{3\!}S_{1}\right)}$.
The Gaussian distribution (gray, green in online version)
and the excess data at the bottom of the figure
(black, red in online version) are explained in the text.
}
\label{mumuAll2S}
\end{figure}

The selection procedure for the data is well described by BABAR
in Refs.~\cite{PRD78p112002,PRL104p191801,ARXIV09100423}.
In Fig.~\ref{mumuAll2S}, we study
the invariant-mass distribution of muon pairs
obtained from the BABAR data set \cite{PRD78p112002}
for the reaction $e^{+}e^{-}$ $\to$
$\Upsilon\left( 2\,{}^{3\!}S_{1}\right)$ $\to$
$\pi^{+}\pi^{-}\Upsilon\left( 1\,{}^{3\!}S_{1}\right)$
$\to$ $\pi^{+}\pi^{-}\mu^{+}\mu^{-}$,
and for a bin size equal to 9 MeV.

Furthermore, we show in Fig.~\ref{mumuAll2S}
a simple Gaussian distribution
(gray histogram, green in online version),
with a width of 89 MeV, around the
$\Upsilon\left( 1\,{}^{3\!}S_{1}\right)$ peak.
We observe that, with respect to the Gaussian distribution,
there is an excess of data
for $M_{\mu^{+}\mu^{-}}$ below
the $\Upsilon\left( 1\,{}^{3\!}S_{1}\right)$ mass,
and a deficit of data
for $M_{\mu^{+}\mu^{-}}$ thereabove.
Actually, we have chosen the Gaussian distribution such
that the total difference between
the data under the Gaussian histogram
and the experimental data vanishes.
The excess signal is indicated
(dark, shaded, histogram, red in online version)
at the bottom of Fig.~\ref{mumuAll2S}.

We observe from Fig.~\ref{mumuAll2S}
that the excess of data sets out for masses
some 40 MeV below
the $\Upsilon\left( 1\,{}^{3\!}S_{1}\right)$ mass,
viz.\ at about $M_{\mu^{+}\mu^{-}}=9.42$ GeV, and then
towards lower $\mu^{+}\mu^{-}$ injvariant masses,
leaving a small signal on top of the increasing background tail,
up to about 9.33 GeV.
The deficit data exhibit enhancements at about
$M_{\mu^{+}\mu^{-}}=9.50$, 9.54 and 9.57 GeV,
i.e., 38, 76, and 114 MeV above
the $\Upsilon\left( 1\,{}^{3\!}S_{1}\right)$ mass, respectively.

\begin{figure}[htpb]
\begin{center}
\begin{tabular}{c}
\raisebox{10pt}{\bf a}\includegraphics[width=230pt]{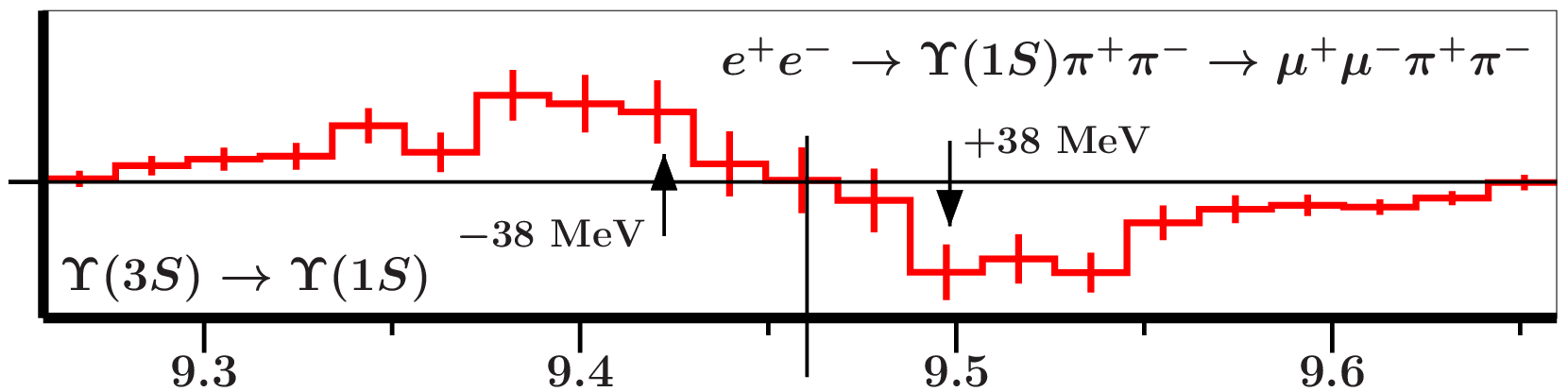}\\
\raisebox{10pt}{\bf b}\includegraphics[width=230pt]{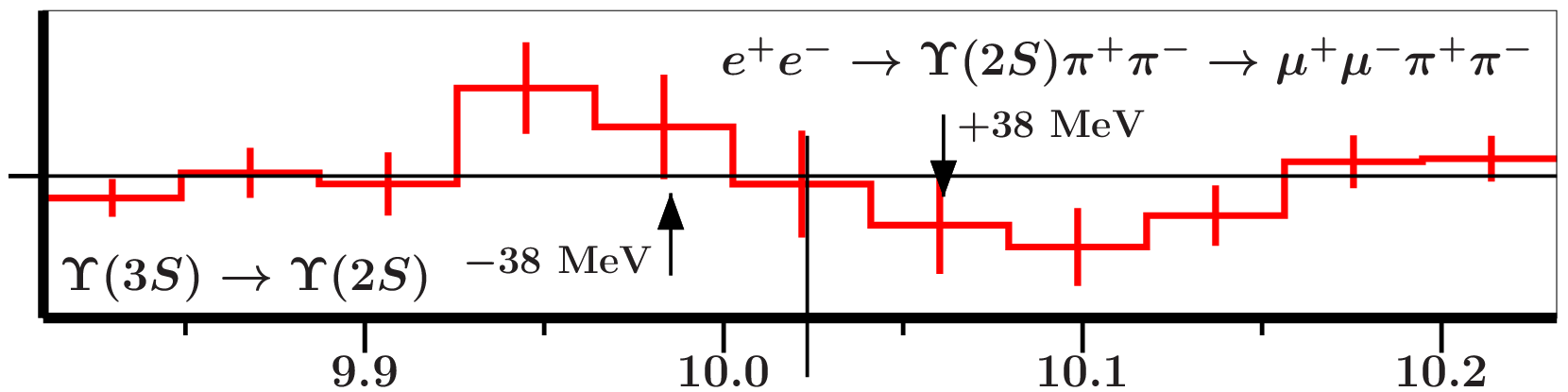}\\
\raisebox{10pt}{\bf c}\includegraphics[width=230pt]{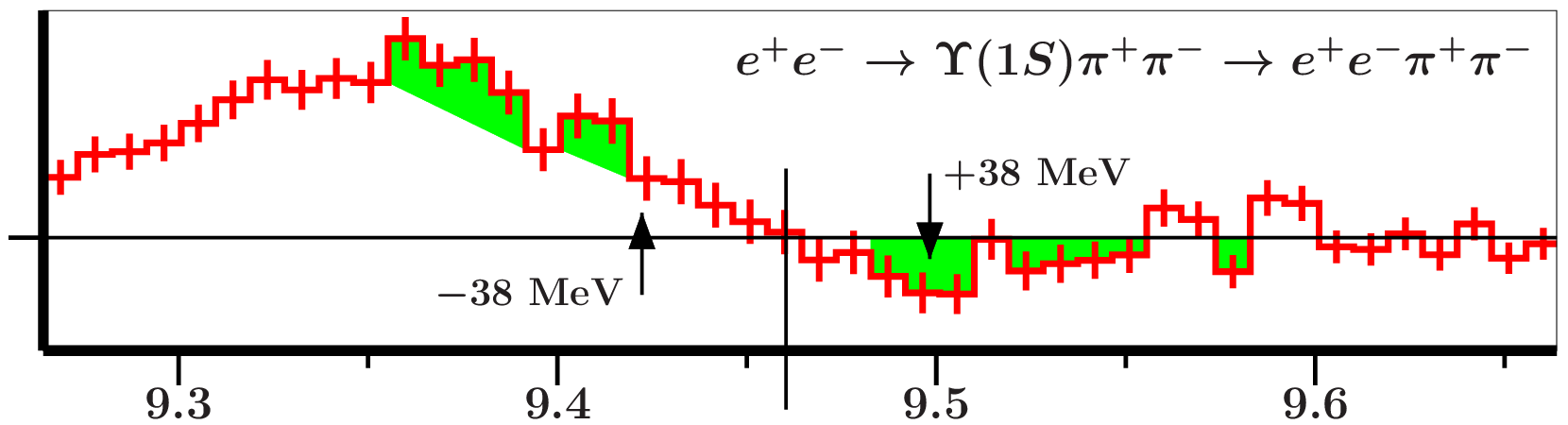}\\ [-10pt]
\end{tabular}
\end{center}
\caption{\small
Excess signals
in invariant $\mu^{+}\mu^{-}$ (a, b)
and $e^{+}e^{-}$ (c) mass distributions:
for the reactions
$\Upsilon\left( 3\,{}^{3\!}S_{1}\right)$ $\to$
$\pi^{+}\pi^{-}\Upsilon\left( 1\,{}^{3\!}S_{1}\right)$
$\to$ $\pi^{+}\pi^{-}\mu^{+}\mu^{-}$,
using bins of 19 MeV (a);
$\Upsilon\left( 3\,{}^{3\!}S_{1}\right)$ $\to$
$\pi^{+}\pi^{-}\Upsilon\left( 2\,{}^{3\!}S_{1}\right)$
$\to$ $\pi^{+}\pi^{-}\mu^{+}\mu^{-}$,
using bins of 38 MeV (b);
$e^{+}e^{-}$ $\to$
$\pi^{+}\pi^{-}\Upsilon\left( 1\,{}^{3\!}S_{1}\right)$
$\to$ $\pi^{+}\pi^{-}e^{+}e^{-}$
for all available data,
using bins of 9 MeV (c).
Statistical errors are shown by vertical bars.
The vertical lines indicate
$M_{\mu^{+}\mu^{-}}=M_{\Upsilon\left( 1,2\,{}^{3\!}S_{1}\right)}$.
Horizontal units,
for $M_{\mu^{+}\mu^{-}}$ (a, b)
and $M_{e^{+}e^{-}}$ (c),
are given in GeV.
}
\label{alldiff}
\end{figure}

In Fig.~\ref{alldiff},
we have collected excess signals for other reactions,
thereby following similar procedures as before.
We have selected all reactions
with some reasonable statistics
from BABAR \cite{PRD78p112002} data,
viz.\
$\Upsilon\left( 3\,{}^{3\!}S_{1}\right)$ $\to$
$\pi^{+}\pi^{-}\Upsilon\left( 1\,{}^{3\!}S_{1}\right)$
$\to$ $\pi^{+}\pi^{-}\mu^{+}\mu^{-}$
(Fig.~\ref{alldiff}a),
$\Upsilon\left( 3\,{}^{3\!}S_{1}\right)$ $\to$
$\pi^{+}\pi^{-}\Upsilon\left( 2\,{}^{3\!}S_{1}\right)$
$\to$ $\pi^{+}\pi^{-}\mu^{+}\mu^{-}$
(Fig.~\ref{alldiff}b),
and
$e^{+}e^{-}$ $\to$
$\pi^{+}\pi^{-}\Upsilon\left( 1\,{}^{3\!}S_{1}\right)$
$\to$ $\pi^{+}\pi^{-}e^{+}e^{-}$
for all available data
(Fig.~\ref{alldiff}c).
The data binning has been chosen
in order to optimize statistics.

In Fig.~\ref{alldiff}a, which is 19 MeV binned,
we observe two connected enhancements at 38 and 76 MeV
below the $\Upsilon\left( 1\,{}^{3\!}S_{1}\right)$ mass,
and a third one, 38 MeV further downwards.
Above the $\Upsilon\left( 1\,{}^{3\!}S_{1}\right)$ mass,
we observe two connected negative enhancements,
38 and 76 MeV higher up in mass.
In Fig.~\ref{alldiff}b, which is 38 MeV binned,
we observe two connected enhancements at 38 and 76 MeV
below the $\Upsilon\left( 2\,{}^{3\!}S_{1}\right)$ mass,
and three connected enhancements at 38, 76 and 114 MeV
above the $\Upsilon\left( 2\,{}^{3\!}S_{1}\right)$ mass.

In itself,
it is not surprising that an intrinsic asymmetry
in mass distributions leads to excess on one side of the center
and to a deficit on the other side,
with respect to a symmetric distribution.
However, we do observe structure in the excess and deficit data.
This can be most clearly seen in the excess distribution
of the reaction $e^{+}e^{-}$ $\to$
$\pi^{+}\pi^{-}\Upsilon\left( 1\,{}^{3\!}S_{1}\right)$
$\to$ $\pi^{+}\pi^{-}e^{+}e^{-}$ (see Fig.~\ref{alldiff}c),
where we do not opt for an overall vanishing excess,
as we did for the other reactions.
The excess signal below
$M_{e^{+}e^{-}}=M_{\Upsilon\left( 1\,{}^{3\!}S_{1}\right)}$
is mainly due to Bremsstrahlung,
as explained by BABAR in Ref.~\cite{PRD78p112002}.
Nevertheless, on top of the Bremsstrahlung background
one observes something extra in the invariant-mass interval
9.35--9.42 GeV.
No doubt, it is an additional signal of hardly more than 1$\sigma$.
However, it is in roughly the same invariant-mass interval
where we find some excess signal in the other three reactions.
Moreover, the pronounced deficit at about 9.50 GeV
comes out also in the same invariant-mass interval
where we find a deficit signal in the other three reactions.
The deficit enhancements at around 9.54 and 9.575 GeV
are hardly distinguishable from zero,
but show up in the expected energy intervals
of 76 and 114 MeV above
the $\Upsilon\left( 1\,{}^{3\!}S_{1}\right)$ mass.

\begin{figure}[htpb]
\begin{center}
\begin{tabular}{c}
\includegraphics[width=240pt]{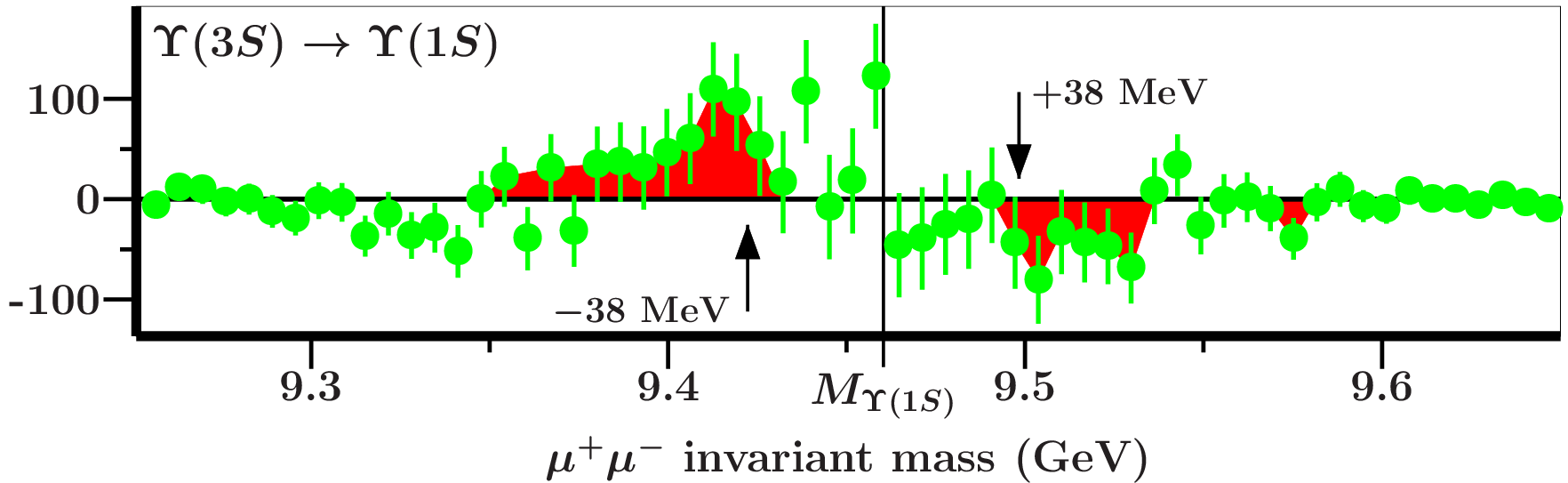}\\ [-10pt]
\end{tabular}
\end{center}
\caption{\small
Event distribution of the excess signal
taken from Ref.~\cite{ARXIV09100423},
in the invariant-$\mu^{+}\mu^{-}$-mass distribution
for the reaction
$\Upsilon\left( 2\,{}^{3\!}S_{1}\right)$ $\to$
$\pi^{+}\pi^{-}\Upsilon\left( 1\,{}^{3\!}S_{1}\right)$
$\to$ $\pi^{+}\pi^{-}\mu^{+}\mu^{-}$,
using bins of 6.5 MeV.
Statistical errors are shown by vertical bars.
The shaded areas (dark, red in online version)
are discussed in the text.
The vertical line indicates
$M_{\mu^{+}\mu^{-}}=M_{\Upsilon\left( 1\,{}^{3\!}S_{1}\right)}$.
}
\label{elisa}
\end{figure}

We found our procedure confirmed
in Ref.~\cite{ARXIV09100423},
where the asymmetry was studied by the BABAR Collaboration itself,
though based on a much larger set of data than here at our disposal.
Their result is shown in Fig.~\ref{elisa}.
Moreover, the analysis in Ref.~\cite{ARXIV09100423}
took all known possible origins of asymmetry into account.
Consequently, what is left (see Fig.~\ref{elisa})
cannot be explained by known physics.

Moreover, Ref.~\cite{ARXIV09100423} confirms
our assumption that for $\mu^{+}\mu^{-}$ background is small.
Also, it states that the, here reported,
systematic uncertainties due to the differences
between data and simulation in the processes
$\Upsilon\left( 1\,{}^{3\!}S_{1}\right)$ $\to$ $\tau^{+}\tau^{-}$
and
$\Upsilon\left( 1\,{}^{3\!}S_{1}\right)$ $\to$ $\mu^{+}\mu^{-}$
cancel, at least in part, in their ratio. This
implies that a similar excess is found
in the
$\Upsilon\left( 1\,{}^{3\!}S_{1}\right)$ $\to$ $\tau^{+}\tau^{-}$
decay.

In order to explain the structures in the deficit signal,
we must assume that the $E(38)$ can be loosely bound
inside a $b\bar{b}$ state, giving rise to a kind of hybrid.
This was discussed to some detail in Ref.~\cite{ARXIV11021863}.

\begin{figure}[htpb]
\begin{center}
\begin{tabular}{c}
\includegraphics[width=230pt]{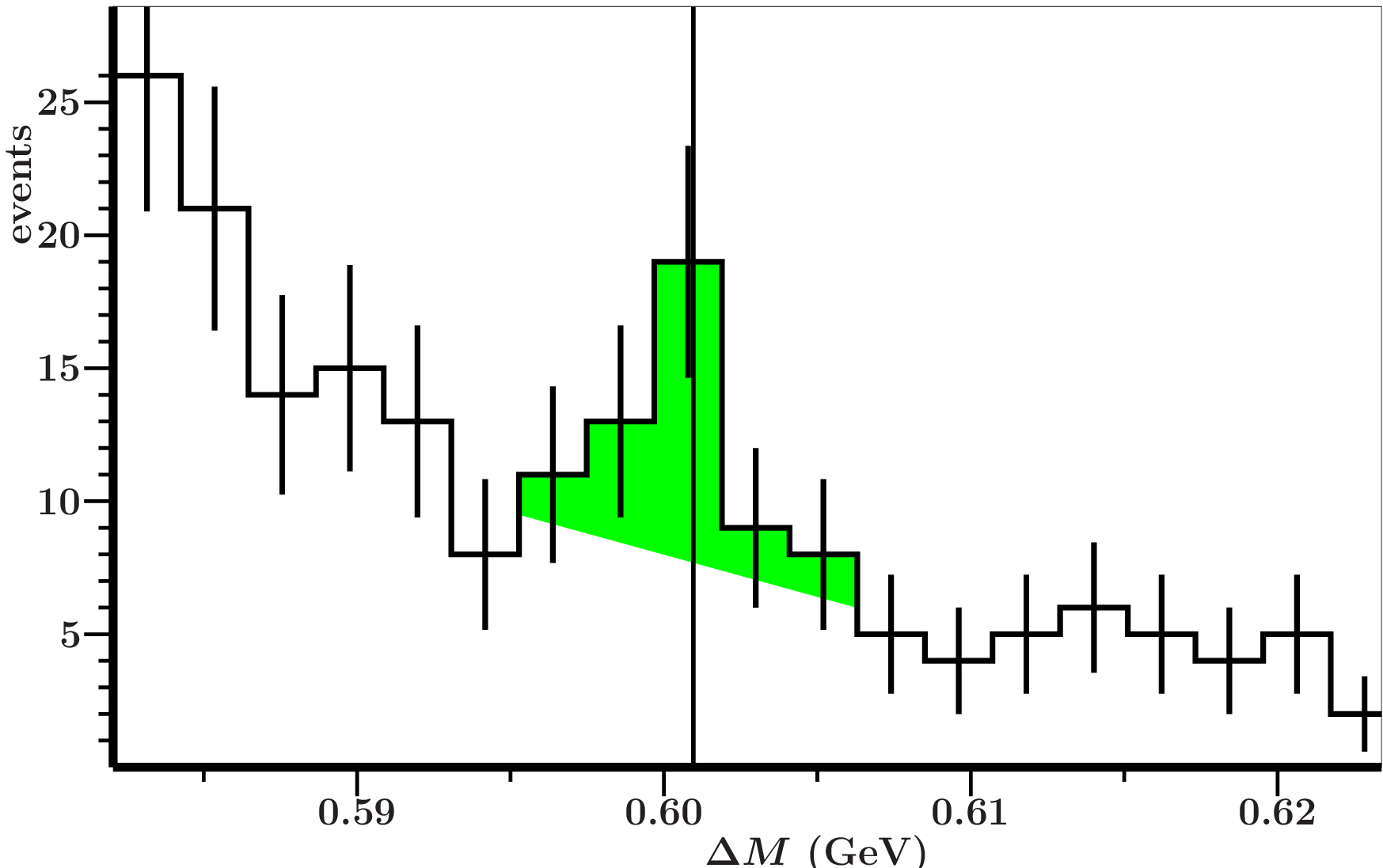}\\ [-10pt]
\end{tabular}
\end{center}
\caption{\small
Possible sign of the $\Upsilon '\left( 2\,{}^{3\!}S_{1}\right)$ hybrid.
The vertical line indicates where
$M_{\Upsilon\left( 2\,{}^{3\!}S_{1}\right)}+38$ MeV comes out,
in terms of $\Delta M=M_{\pi^{+}\pi^{-}\mu^{+}\mu^{-}}-M_{\mu^{+}\mu^{-}}$.
Data are from Ref.~\cite{PRD78p112002}.
}
\label{hybrid}
\end{figure}

In Fig.~\ref{hybrid}, we show the event distribution
for the invariant mass $\Delta M$, which is defined
\cite{PRD78p112002} by $\Delta M=$
$M_{\pi^{+}\pi^{-}\mu^{+}\mu^{-}}-M_{\mu^{+}\mu^{-}}$,
where the latter mass is supposed
to be the $\Upsilon\left( 1\,{}^{3\!}S_{1}\right)$ mass.
Thus, a signal with the shape of a narrow Breit-Wigner resonance
seems to be visible on the slope of the
$\Upsilon\left( 2\,{}^{3\!}S_{1}\right)$
resonance, though with little more than 2$\sigma$ relevance.
Nevertheless, by coincidence or not, it comes out exactly in the
expected place, namely at
$M_{\Upsilon\left( 2\,{}^{3\!}S_{1}\right)}+38$ MeV.

New evidence of the existence of $E(38)$
comes from observations published
by the CB-ELSA Collaboration \cite{EPJA33p133}
and by the COMPASS Collaboration \cite{ARXIV11090272,ARXIV11086191}.

In Ref.~\cite{EPJA33p133} the CB-ELSA Collaboration
studied photoproduction of $\eta$-mesons off protons
with the Crystal-Barrel detector at ELSA, for photon
energies in the range from 0.75 to 3 GeV.
Their data were taken in three run periods, more than a decade ago,
with electron-beam energies of 1.4, 2.6, and 3.2 GeV.
Of our interest here are, in particular, $\eta$-mesons
detected in $\eta\to\gamma\gamma$,
for which $\gamma\gamma$ cross sections were presented.

\begin{figure}[htpb]
\begin{center}
\begin{tabular}{c}
\includegraphics[width=230pt]{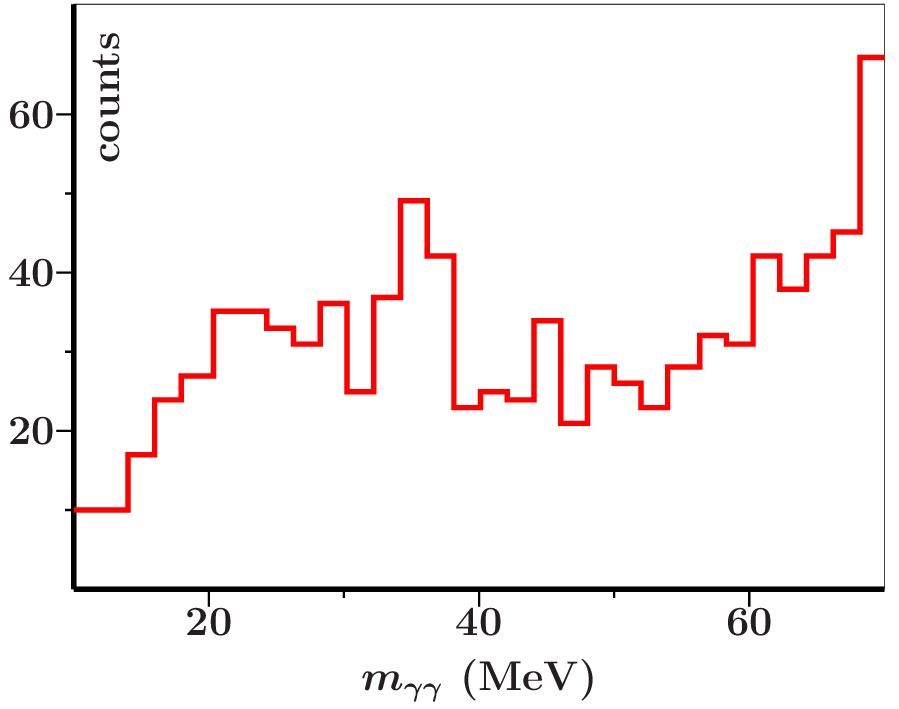}\\ [-10pt]
\end{tabular}
\end{center}
\caption{\small
A modest signal at about 37 MeV in the two-photon distribution
published by the CB-ELSA Collaboration \cite{EPJA33p133}.}
\label{cbelsagammagamma}
\end{figure}

In Fig.~\ref{cbelsagammagamma} we show a detail
of the two-photon invariant-mass distribution
published by the CB-ELSA Collaboration
\cite{EPJA33p133} for the reactions $p\gamma\to p\gamma\gamma$
and $p\gamma\to p3\pi^{0}$, extracted from the 3.2 GeV data.
Although with low statistics, one can observe a narrow peak in the CB-ELSA
data centered at about 37 MeV.
Clearly, the signal is too small to claim evidence of $E(38)$,
but it illustrates the possible decay mode $E(38)\to\gamma\gamma$
through loops of virtual light quarks.

We do not expect the $E(38)\to\gamma\gamma$ mode
to be very large, since $E(38)$ couples to quarks proportionally to their
masses, as we concluded above in connection with the observed oscillations.
Nevertheless, the two-photon data of the CB-ELSA Collaboration
stimulate us to search for similar data, in other experiments, supporting
$E(38)$.

\begin{figure}[htpb]
\begin{center}
\begin{tabular}{c}
\includegraphics[width=230pt]{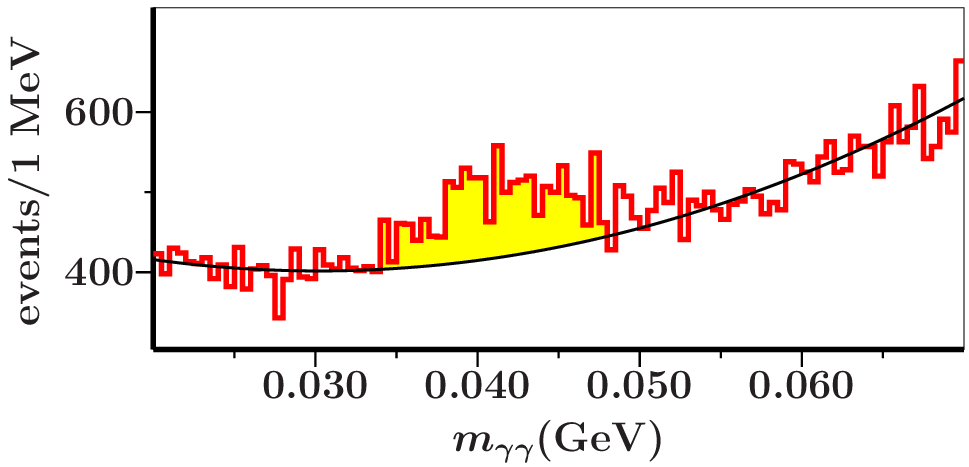}\\ [-10pt]
\end{tabular}
\end{center}
\caption{\small
A modest signal in the $\gamma\gamma$ COMPASS \cite{ARXIV11090272} data
at around 40 MeV.}
\label{bewaar}
\end{figure}

In Ref.~\cite{ARXIV11090272}, the COMPASS Collaboration
studied $\omega$ and $\phi$ vector meson production in
$pp\to pp\omega /\phi$ data,
obtained at the two-stage magnetic COMPASS spectrometer
attached to the SPS accelerator facility at CERN.
For the indentification of the $\omega$ meson, the $\pi^{0}$
has been reconstructed from two photons.
A detail of the thus obtained invariant-mass distribution
for $\gamma\gamma$ pairs is shown in Fig.~\ref{bewaar}
in which an enhancement at about 40 MeV can be observed.

In Ref.~\cite{ARXIV11086191}, the COMPASS Collaboration
carried out a partial-wave analysis of $p\pi^{-}\to p\pi^{-}\eta '$
in order to extract the exotic $J^{PC}=1^{-+}$ $\pi^{-}\eta '$ $P$-wave
signal. The analysis consists of various intermediate steps.
The first step is to select $\pi^{0}\eta$ pairs,
possibly stemming from the decay of an $\eta'$-meson.
Either one of the two mesons $\pi^0,\eta$ or both may decay into pairs of
photons. Hence, as a byproduct of their analysis,
the COMPASS Collaboration produced an invariant-mass distribution
of photon pairs.

\begin{figure}[htpb]
\begin{center}
\begin{tabular}{c}
\includegraphics[width=230pt]{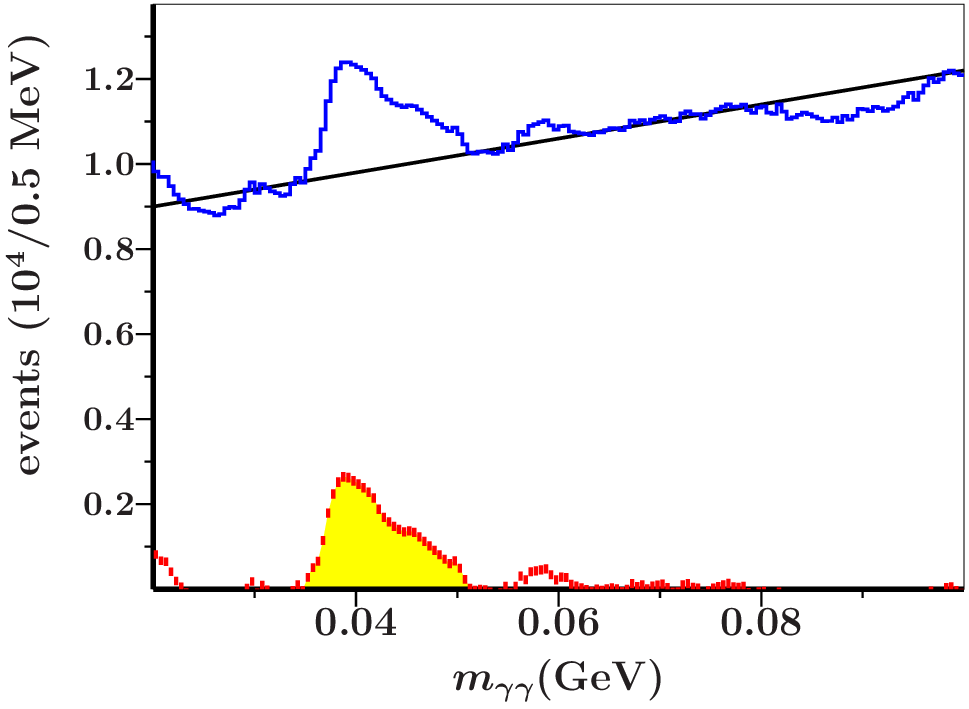}\\ [-10pt]
\end{tabular}
\end{center}
\caption{\small
Top: a clear signal in the $\gamma\gamma$ COMPASS \cite{ARXIV11086191} data,
with maximum at about 39 MeV. Bottom: the $E(38)$ structure
that remains after background subtraction and contains about 46000 events.}
\label{compassgammagamma}
\end{figure}

In Fig.~\ref{compassgammagamma} we show a detail of
the invariant two-photon mass distribution of COMPASS
\cite{ARXIV11086191}.
These data seem to have enough statistics to substantiate the existence of
a light boson with mass around 40 MeV.

With this latest piece of evidence of $E(38)$,
we conclude that it is now necessary to establish its mass
and other properties by further experiments.
We think that $E(38)$ is the light scalar Higgs-type boson
which was proposed in a model describing the unification
of electromagnetic and strong interactions \cite{NCA80p401}.
Furthermore, we believe that it consists of a micro-universe
filled with glue, as formulated in
Refs.~\cite{JMP27p1411,PRD30p1103}.

Finally, as $E(38)$ appears to couple to quarks
proportionally to their masses,
its coupling to the top quark is expected to be quite strong.

\section*{Acknowledgments}

We are grateful for the precise measurements
and data analyses of the BABAR, CDF, CMD-2,
CB-ELSA, and COMPASS Collaborations,
which made the present analysis possible.
One of us (EvB) wishes to thank
Drs. B.~Hiller, A.~H.~Blin, and A.~A.~Osipov
for useful discussions.
This work was supported in part by the {\it Funda\c{c}\~{a}o para a
Ci\^{e}ncia e a Tecnologia} \/of the {\it Minist\'{e}rio da Ci\^{e}ncia,
Tecnologia e Ensino Superior} \/of Portugal, under contract no.\
CERN/\-FP/\-116333/\-2010.

\newcommand{\pubprt}[4]{#1 {\bf #2}, #3 (#4)}
\newcommand{\ertbid}[4]{[Erratum-ibid.~#1 {\bf #2}, #3 (#4)]}
\def\EPJA{Eur.\ Phys.\ J.\ A}
\def\JETPL{JETP Lett.}
\def\JMP{J.\ Math.\ Phys.}
\def\NCA{Nuovo Cim.\ A}
\def\PRD{Phys.\ Rev.\ D}
\def\PRL{Phys.\ Rev.\ Lett.}

\end{document}